\documentclass[12pt]{article}
\usepackage{graphicx}  
\usepackage{pdfpages}
\usepackage{amsmath}   
\usepackage{amssymb}   
\usepackage{bm} 
\usepackage{dcolumn}
\usepackage{color}
\usepackage{mathrsfs}
\usepackage{amsfonts}
\usepackage{varioref}
\usepackage[utf8]{inputenc}
\usepackage{amsmath}
\usepackage{amsfonts}
\usepackage{amssymb}
\usepackage{enumitem}
\usepackage{float}
\usepackage{epsfig}
\usepackage{amsfonts}
\usepackage{latexsym}
\usepackage{hyperref}
\usepackage{setspace}
\usepackage{bm}
\usepackage{slashed}
\usepackage{graphicx}
\usepackage{bm}
\usepackage{hyperref}
\usepackage{dsfont}
\usepackage{caption}
\usepackage{subcaption}
\def\be{\begin{equation}}
\def\ee{\end{equation}}
\labelformat{section}{Section #1}
\labelformat{subsection}{Section #1}
\labelformat{subsubsection}{Section #1}
\labelformat{subsubsubsection}{Section #1}
\labelformat{equation}{Eq.~(#1)}
\labelformat{figure}{Fig.~#1}
\labelformat{subfigure}{Fig.~\thefigure#1}
\labelformat{table}{Tab.~#1}
\labelformat{appendix}{Appendix #1}
\title{\bf Effect of back reaction on entanglement and subregion volume complexity in strongly coupled plasma. }
\author{Shankhadeep Chakrabortty\footnote{s.chakrabortty@iitrpr.ac.in}, ~~Sanjay Pant\footnote{2018phz0012@iitrpr.ac.in} ~~and ~Karunava Sil\footnote{karunava.sil@iitrpr.ac.in}\\
{{\small  Department of Physics, Indian Institute of Technology Ropar}}\\ {{\small Rupnagar, Punjab 140 001,
India}}}

\begin{document}
\maketitle
\begin{abstract}
\noindent The back reaction imparted by a uniform distribution of heavy static fundamental quarks on  large $N_c$ strongly coupled  gauge theory can be holographically realized as a deformation in $AdS$ blackhole background. The presence of back reaction brings  significant changes in to the entanglement structure of the strongly coupled boundary theory at finite temperature. Since the deformed blackhole geometry still remains asymptotically $AdS$, the $gauge/gravity$ duality allows us to explore the entanglement structure of back reacted plasma in a quantitative way by computing various measures, e.g holographic entanglement entropy (HEE) and entanglement wedge cross section (EWCS). We explicitly study the variation of those entanglement measures with respect to the uniform density of heavy static fundamental quarks present in the boundary theory. In particular, we notice enhancement of both HEE and EWCS with respect to quark density. We also study the effect of back reaction on the holographic subregion volume complexity. In this analysis we observe an occurrence of logarithmic divergence proportional to the quark density parameter.

\end{abstract}
\vskip .2cm
\noindent
\bigskip
\section{Introduction}

Quantizing gravity by following the standard rule of local quantum field theory encounters many troubles. Several studies on this issue indicate that the microscopic degrees of freedom of gravity is fundamentally different in nature as compared to that of other fundamental interactions.
 A strong evidence for adopting such idea originates from the Bekenstein-Hawking (BH) entropy of blackhole \cite{Bekenstein, Hawking1} that quantifies the microscopic data of the blackhole spacetime and most importantly turns out to be proportional to the surface area of the blackhole event horizon whereas in all other non-gravitational theories, entropy is proportional to the volume. Such unusual properties of gravitational degrees of freedom leads to a proposal called \textit{holographic principle} that suggests as unlike the case of usual field theory, the information of microscopic degrees of freedom contained in a $d+1$ dimensional gravitational theory is not proportional to the volume of the spacetime but to the area of a $d$ dimensional boundary enclosing that space time \cite{Stephens:1993an, Susskind:1994vu}.  A concrete example of holographic principle discovered in string theory is known as $AdS/CFT$ correspondence that relates a quantum theory of gravity in $d+1$ dimensional spacetime with negative cosmological constant to a non-gravitational theory with conformal invariance living on the $d$ dimensional boundary of that spacetime \cite{Maldacena:1997re, Witten:1998qj, Gubser:1998bc, Aharony:1999ti}. There has been a further generalization known as $gauge/gravity$ correspondence which is mostly made out of phenomenological perspective (see \cite{Erdmenger:2018xqz} and references therein). By now, it has been highly admitted that the quantum information theory plays a key role to understand such holographic nature of gravity. In favor of this connection, there has been several proposals that relate various measures of entanglement in the boundary theory to certain kind of geometrical quantities in the dual bulk spacetime.

A lot of progress has been made to build up a good knowledge of various measures of entanglement in non-relativistic quantum information theory \cite{Chuang}. The most well-studied measure of bipartite entanglement is known as entanglement entropy (EE) which is originally defined as the von Neumann entropy of the bipartite system. EE measures the nonlocal correlations between two entangled systems by quantifying the loss of information as one of the systems becomes inaccessible to the observer. EE satisfies a number of important physical properties including the area law and the strong sub-additivity condition. To explore the relation between quantum information and holography, an appropriate generalization of such entanglement measures in quantum field theory (QFT) was very much required. For a general $d+1$ dimensional QFT, the computation of EE goes beyond the scope of analytical method. However, in $1+1$ dimensional conformal field theory (CFT), EE is computed for single interval in an infinite system using the replica trick method and it turns out to be proportional to the central charge of the theory \cite{Calabrese:2004eu, Calabrese:2009qy}. Same analysis holds for 1+1 mass deformed CFT. Moreover, the computation of EE in a finite system as well as for multiple disjoint intervals is also accomplished and a formal extension of various results to higher dimensional CFT has also been proposed \cite{Calabrese:2010he, Calabrese:2009ez}.

Extending the idea of Bekenstein-Hawking entropy, Ryu and Takayanagi (RT) \cite{Ryu:2006bv} conjecture a holographic prescription that establishes a direct relation between the entanglement entropy in the boundary field theory and a geometric quantity, i.e. the area of a spatial minimal surface in the dual bulk spacetime. The holographic formula given by Ryu-Takayanagi has been generalized in a covariant manner in \cite{Hubeny:2007xt} to derive the time dependence of EE, while the correction due to the higher derivative terms in the Lagrangian of gravitational action is given in \cite{Dong:2013qoa}. Holographic EE and mutual information of an infinite strip in large $N$ gauge theory having finite temperature and also finite chemical potential has been proposed in \cite{Fischler:2012ca, Kundu:2016dyk, Fischler:2012uv}. Further, The quantum correction to the holographic entanglement entropy was first introduced in \cite{Faulkner:2013ana}. By imposing a large central charge limit in the holographic CFT, a proof of RT conjecture of holographic entanglement entropy for multiple disjoint intervals is given in \cite{Hartman:2013mia}.

EE is a suitable measure to determine the strength of entanglement of a pure bipartite state. For mixed bipartite state (e.g. thermal state), if we compute the von Neumann entropy by using the holographic RT method, the final result contains both thermal entropy as a leading divergent term and also the entanglement entropy as sub-leading finite term. For a large $N$ gauge theory at finite temperature and at finite chemical potential, such observation on EE is already mentioned in \cite{Kundu:2016dyk}. Therefore, it is desired to analyze some holographic measure other than entanglement entropy which could exclusively estimate the mixed entanglement in a QFT. Recently, a new geometric quantity, the entanglement wedge cross section $(EWCS)$ \cite{Takayanagi:2017knl} has been proposed to be holographic dual of entanglement of purification (EoP) which is known to be a good measure of mixed entanglement in the quantum information theory \cite{Terhal}.
To define EoP, let us consider a bipartite system in a given mixed state $\rho_{AB}$ such that $\rho_{AB}\in \mathcal{H}_{A}\otimes \mathcal{H}_{B}$, with $\mathcal{H}_{A}$ and $\mathcal{H}_{B}$ being the Hilbert spaces for the subsystems $A$ and $B$ respectively. In order to purify this mixed state one needs to enlarge the original Hilbert space by considering additional degrees of freedom, so that the purification takes the following form, $\rho_{AB}=Tr_{\bar{A}\bar{B}}|\psi><\psi|$, where $|\psi> \in \mathcal{H}_{A\bar{A}}\otimes \mathcal{H}_{B\bar{B}}$. However this way of constructing a purification is not unique and the EoP is defined as von Neumann entropy for minimal entanglement between the subregions $A$ and $B$,
\begin{equation}
E_{p}(\rho_{AB})=\mathop{min}_{\rho_{AB}=Tr_{\bar{A}\bar{B}}|{\psi}\rangle \langle{\psi|}}S_{\rho_{A\bar{A}}}.
\end{equation}
In \cite{Takayanagi:2017knl, Nguyen:2017yqw} the holographic dual to EoP has been conjectured as equal to the area of the minimal cross section inside the entanglement wedge connecting the two subregions, called the entanglement wedge cross section.

It is important to note that this connection between EoP and the EWCS is yet to be proved. Nevertheless, the conjecture is based on several common properties  satisfied by both EoP and the EWCS. As an example, in \cite{Terhal} it has been shown that EoP is constrained by the following inequality involving the mutual information (MI),
\begin{equation}\label{MIEOP}
\frac{I(A,B)}{2}\le E_{p}(\rho_{AB}).
\end{equation}
The same relationship holds true between $EWCS$ and $MI$ in the context of $AdS_3/CFT_2$, thereby establishing the above conjecture  \cite{Takayanagi:2017knl}. Further, holographic analysis has been done in support of the above  inequality between EWCS and MI in non-relativistic theories with hyper-scaling violation \cite{BabaeiVelni:2019pkw}, in confining theories as well as in three-dimensional Chern-Simons matter theory with fundamental flavor \cite{Jokela:2019ebz}.

Connection between quantum information theory and gravity witnesses a major improvement once an equivalence between quantum complexity and space time geometry is proposed.
In this context, a very intriguing question to ask is how to quantify the cost of generating a particular state in the boundary field theory in terms of dual bulk geometry. A plausible explanation to this question was first addressed by Suskind in the context of black hole physics \cite{Susskind:2014rva}.
In free quantum field theories the computation of quantum complexity has been performed by adopting a geometric approach due to Nielson \cite{Nielsen}. By definition, complexity counts the minimum number of unitary operators/quantum gates to construct a target state $|{\psi_{T}}\rangle$ starting from a simple reference state $|{\psi_{R}}\rangle$.  The construction of such target state staring from ground state of the free theory, coherent state and also the thermo-field double state has been successfully achieved and also the complexity is computed appropriately \cite{Jefferson:2017sdb, Guo:2018kzl, Chapman:2018hou}. Such analysis turns out to be very difficult to perform in the presence of interaction and it is actually impossible to extend in strongly coupled theories. However in recent time, there has been various proposals for holographic computation of the complexity in the boundary theory. In particular, two independent holographic conjectures are proposed, e.g. the complexity equals volume conjecture or CV-duality and and the complexity equals bulk action or CA-duality. According to the CV-duality, the complexity of the boundary theory is proportional to the volume of co-dimension one hyper surface in the dual bulk geometry \cite{Susskind:2014moa, Susskind:2014jwa}. On the other hand, the statement for the CA-duality suggests a direct relation between complexity and the total action evaluated for the Wheeler-DeWitt patch in the bulk \cite{Brown:2015bva,Brown:2015lvg}. Motivated by these, another proposal for computing complexity of any subregion in the boundary has been proposed as the subregion volume complexity and it measures the complexity of a mixed quantum state in the boundary theory on a constant time slice in terms of the volume of the entanglement wedge inside the RT surface \cite{Alishahiha:2015rta, Ben-Ami:2016qex}. The mixed quantum state can be obtained by reducing the total system on a given pure state to a particular subregions on the boundary. A generalization of subregion volume complexity in order to include the time dependent cases can be found in \cite{Carmi:2016wjl}. In this analysis, apart from the power law divergence the authors also obtain a new logarithmic divergence in the result for complexity using the CA-duality.

It has been quite certain that the $gauge/gravity$ duality offers simplistic ways to understand quantum information theory from the perspective of QFT and it also offers a better understanding of the relation between the information theory and the holographic nature of gravity. The holographic prescription for computing EE, EWCS and also quantum complexity are remarkably simple but thoroughly insightful. The analysis of such useful measures of entanglement turns out to be even more interesting if we consider different modifications on the boundary field theory to make it more realistic. Due to $gauge/gravity$ duality these modifications affects the dual bulk geometry significantly and consequently holographic calculation of different entanglement measures receives non trivial corrections. Examples of such modification includes addition of massive relevant operator to the free theory, deformation of shape for the entangling region in the boundary or addition of some extra degrees of freedom apart from the usual matter content of the boundary theory.

To emphasize the effect of deformation/back reaction on the entanglement structure in the boundary theory we review few earlier works.
Following RT method, holographic EE for a CFT with planar defect has been first discussed in \cite{Jensen:2013lxa}.  In another example, EE has been calculated for spherical entangling surface by considering similar kind of planner defects in super conformal theory where the deformation in the CFT side can be thought of as dual to $M2$ and $M5$ probe branes in the bulk \cite{Rodgers:2018mvq}. Similar analysis has been done for mass deformed field theory and as a consequence a new universal term which diverges as the logarithm of the UV cut-off has been obtained \cite{Hung:2011ta}. In \cite{Kontoudi:2013rla}, considering massive back reacted flavor in SYM theory as realized by the $D3/D7$ brane set up in the bulk, EE is evaluated for both slab and ball shape entangling surface by doing a perturbative expansion in $N_{f}/N_{c}$.
The correction to entanglement entropy due to shape modification of the entangling surface was obtained in \cite{Carmi:2017ezk, Carmi:2015dla, Fonda:2015nma}. Also in \cite{Carmi:2017ezk} considering relevant deformation of the boundary field theory, a computation of subregion volume complexity has been done which reports an universal UV divergent term in the final result for complexity.

It is important to note that although the effect of back reaction on EE and complexity in boundary theory has been analyzed in few occasions, to the best of our knowledge, there has been hardly any work available in the literature that talks about what happens to EoP in the presence of back reaction. This serves as a prime motivation to carry forward this particular project. In this work, we consider $d$ dimensional strongly coupled large $N_c$ gauge theory at finite temperature in presence of a uniform distribution of large number $N_f$ of externally added heavy flavor quarks. The back reaction imparted to the gauge theory by the external heavy quarks is significant if $N_f \sim N_c^2$ or more than that. Within the context of $gauge/gravity$ duality, the dual description to back reacting plasma with quark density consists of a uniform distribution of string cloud such that one end of each string is attached to the boundary while the string itself is extended deep into the bulk along the radial direction. Here we have considered the string cloud distribution to be homogenous and ignored any interaction between them, however their back reaction to the geometry was included in the bulk metric resulting a $d+1$ dimensional deformed AdS blackhole geometry \cite{Chakrabortty:2011sp, Chakrabortty:2016xcb}. With this background we holographically study the entanglement entropy, entanglement of purification and quantum complexity of strongly coupled large $N_c$ gauge theory at finite temperature and attempt to capture the effect of back reaction due to the presence of external heavy flavor quark on each of them. It is very important to note that the gauge/gravity duality offers technics to identify various universal hydrodynamical properties of strongly coupled holographic plasma. Most importantly, such universal properties show qualitative agreement at the experimental level with strongly coupled QGP medium \cite{Arnold:2007pg, Shuryak:2008eq, Policastro:2001yc}. Since the hydrodynamical descriptions of both strongly coupled holographic plasma and QGP rely on a underlying microscopic structure, it is very relevant question to ask what happens if mixed bipartite entanglement structure prevails in those microscopic theories. Within the framework of $gauge/gravity$ duality our present paper attempts a systematic study to address this issue. Moreover, we emphasize the effect of back reaction in the strongly coupled holographic plasma in the context of mixed bipartite entanglement which, we hope, may shed some light to understand the mixed bipartite entanglement structure in the QGP in the presence of other heavy quarks.

The organization of the paper is as follows. In section $2$ we briefly discuss about the back reacted strongly coupled plasma and its gravity dual. In section $3$, we present the computation of holographic entanglement entropy. The details of  EWCS is discussed in section $4$. Further, we elaborate upon subregion complexity in section $5$. Finally we conclude in section $6$.

\section{Gravitational background dual to quark cloud model}\label{background}
After the pioneering work \cite{Karch:2002sh}, there has been a lot of attempts to study the characteristics of strongly coupled gauge theory by using the probe approximation. However, going beyond probe approximation is hardly achievable except very few cases as the appropriate gravity background dual to the back reacted boundary theory is very hard to compute. One of the authors of the present work has been able to construct a gravitational background which is dual to the strongly coupled large $N_c$ gauge theory back reacted by the presence of a uniform distribution of external heavy quarks \cite{Chakrabortty:2011sp}.  As previously mentioned, the quark degrees of freedom on the boundary are dual to the the homogeneous distribution of strings in the bulk producing a nontrivial deformation of the AdS-BH metric. The $d+1$ dimensional gravitational action is given as \cite{Chakrabortty:2011sp},
\begin{equation}
\label{action}
S=\frac{1}{4\pi G_{d+1}}\int dx^{d+1}\sqrt{g}\left(R-2\Lambda\right)+S_{M},
\end{equation}
where, $S_{M}$ stands for the matter part of the action arising due to the presence of uniform distribution of strings,
\begin{equation}
\label{matteraction}
S_{M}=-\frac{1}{2}\sum_{i}\mathcal{T}_{i}\int d^2\xi \sqrt{-h}h^{\alpha\beta}\partial_{\alpha}X^{\mu}\partial_{\beta}X^{\nu}g_{\mu\nu}.
\end{equation}
In the above action, $g_{\mu\nu}$ represents the target spacetime and $h_{\alpha\beta}$ is the intrinsic metric of the string world-sheet and $\mathcal{T}_{i}$ representing the tension of the $i$th string.
Varying  this action with respect to the space-time metric leads to
\begin{equation}
R_{\mu\nu}-\frac{1}{2}R g_{\mu\nu}+\Lambda g_{\mu\nu}=8\pi G_{\mu\nu}T_{\mu\nu}
\end{equation}
with,
\begin{equation}
T^{\mu\nu}=-\sum_{i}\mathcal{T}_{i}\int{d^2\xi\frac{1}{\sqrt{\mid{g_{\mu\nu}}}}\sqrt{-h^{\alpha\beta}}h^{\alpha\beta}\partial_{\alpha}X^{\mu}\partial_{\beta}X^{\nu}g_{\mu\nu}\delta_{i}^{d-1}(x-X_i)}
\end{equation}
The density function repressing the distribution of the uniform string cloud is given as,
\begin{equation}
\label{metric}
\nonumber b(x)=\mathcal{T}\sum_{i=1}^{N}\delta_{i}^{(d-1)}(x-X_{i}),
\end{equation}
where it is assumed that the tension for each of the strings are equal to $\mathcal{T}$ with $N$ being the total number of strings. Averaging over the $(d-1)$ spatial dimensions the constant string density can be defined as,
\begin{equation}
\label{metric2}
\nonumber \tilde{b}=\frac{1}{V_{d-1}}\int b(x)d^{d-1}x=\frac{\mathcal{T}N}{V_{d-1}},
\end{equation}
where $V_{d-1}$ is the volume of the $(d-1)$ dimensional space. In the limit $V_{d-1}\rightarrow \infty$, we consider very large value of $N$ to keep $N/V_{d-1}$ finite.
The non vanishing components of $T^{\mu\nu}$ are
\begin{equation}
T_{00}=-\frac{\tilde{b}}{r^3}g_{tt}~~~~~~~~~~~T_{rr}=-\frac{\tilde{b}}{r^3}g_{rr}
\end{equation}
The ansatz for the AdS-BH metric can be written as,
\begin{equation}
ds^2=-V(r)dt^2+\frac{dr^2}{V(r)}+\frac{r^2}{R^2}\delta_{ij}dx^{i}dx^{j},
\end{equation}
with the explicit form of $V(r)$, given as,
\begin{equation}
\label{V}
V(r)=K+\frac{r^2}{R^2}-\frac{2m}{r^{d-2}}-\frac{2b R^{d-3}}{(d-1)r^{d-3}},
\end{equation}
where the value of $K$ is equal to $0,1,-1$ for the $d-1$-dimensional boundary to be flat, spherical or hyperbolic  respectively. In this paper we have decided to work with $K=0$. Moreover, in the above expression the string cloud density is represented by the dimension less quantity $b=\tilde{b}R$.

It turns out to be very useful to write the metric in terms of the radial coordinate $z=\frac{R^2}{r}$ to get,
\begin{equation}
\label{metric3}
ds^2=\frac{R^2}{z^2}\left(-h(z)dt^2+d\vec{x}^2+\frac{dz^2}{h(z)}\right),
\end{equation}
where the spatial coordinate of the boundary is represented by the $d-2$ dimensional vector $\vec{x}$. The function $h(z)$ reads as,
\begin{equation}
\label{h}
h(z)=1-\frac{2m}{R^{2d-2}}z^{d}-\frac{2b}{(d-1)R^{d-1}}z^{d-1}.
\end{equation}

It is useful to write equation (\ref{h}) as the following,
\begin{equation}
h(z)=\left[1-\rho\left(\frac{z}{z_{H}}\right)^{d-1}+(\rho-1)\left(\frac{z}{z_{H}}\right)^d\right],
\end{equation}
where, $\rho$ is a dimensionless quantity defined as
\begin{eqnarray}
\rho =\frac{2bz_{H}^{d-1}}{(d-1)R^{d-1}},
\end{eqnarray}

The Hawking temperature of the deformed AdS black is now expressed as,
\begin{equation}
\label{tem}
T=-\frac{1}{4\pi}\frac{d}{dz}h(z)\Biggl|_{z=z_{H}}=\frac{(d-\rho)}{4\pi z_{H}}.
\end{equation}
In this parametrization the range of $\rho$ is found to be $0\leq \rho\leq d$, where the minimum value corresponds to zero quark density while the maximum defines the zero black hole temperature. The thermal entropy can be calculated as \cite{Chakrabortty:2011sp},
\begin{equation}
S=\int T^{-1}dM,
\end{equation}
$M$ being the ADM mass of the black hole given in terms of the integration constant $m$ as appeared in the above solution as, $M=\frac{(d-1)m V_{d-1}}{8\pi G_{d+1}}$. Moreover the entropy density is given as,
\begin{equation}
s=\frac{R^{2d-2}}{4G_{d+1}}\frac{1}{z_{H}^{d-1}}
\end{equation}
The above result for entropy density suggests the following definition of an effective temperature $T_{f}$,
\begin{equation}
\label{Tf1}
T_{f}=\frac{d}{4\pi z_{H}},
\end{equation}
such that now the thermal entropy density $s$ is proportional to $T_{f}^{d-1}$. Also, as we will see in section $3$ that the low temperature entanglement entropy contains a term which is proportional to $T_{f}^{d-1}$. So it is more appropriate to work with the effective temperature rather than the actual temperature as given in eqn. (\ref{tem}). This particular deformed AdS black hole background is thermodynamically stable as well as geometrically stable up to tensor and vector perturbation \cite{Chakrabortty:2011sp}.
\section{Holographic Entanglement Entropy}
In this section using prescription by Ryu-Takayanagi, we will study the holographic entanglement entropy of a strongly coupled SYM plasma at finite temperature along with an uniform distribution of heavy fundamental quark. Given a bipartite quantum state, entanglement entropy quantifies the amount of entanglement between any subsystem $A$ and it's compliment $B$ on a constant time slice in the $d$ dimensional boundary theory. In general, the subsystem $A$ can be of any shape, here we consider it to be an infinitely long strip with finite width $l$. In the dual gravity side, a hypersurface with minimal area extending into the bulk with boundary same as that of the subsystem $A$ can be constructed. This hypersurface with minimal area is known as the Ryu-Takayanagi surface $\gamma_{A}$. According to the proposal, EE is given by the area of the RT surface divided by the Newton's constant,
\begin{equation}
S_{A}=\frac{\textrm{Area}[\gamma_{A}]}{4G_{N}}.
\end{equation}
In the following we define the strip like entangling region as,

\begin{equation}
\label{strip surface}
 x\in\Biggl[-\frac{l}{2},\frac{l}{2}\Biggr];~~~~~~~y_i\in\Biggl[-\frac{L}{2},\frac{L}{2}\Biggr];~~~~~ i={1,...,d-2.}
\end{equation}
The profile of the extremal hyper surface is defined by the radial dependance of the coordinate $x=x(z)$

and the corresponding area functional to be minimized is given by,
\begin{equation}
\label{HEE}
\mathcal{A}=2L^{d-2}R^{d-1}\int\frac{dz}{z^{d-1}}\sqrt{\left(\frac{dx}{dz}\right)^2+\frac{1}{h(z)}}.
\end{equation}
Note that $x$ is cyclic in the above action (\ref{HEE}) and as a consequence of that one can obtain a first integral of motion to be determined by using the boundary condition, $\lim_{x\to\infty} z=z_t$. Here we define $z_{t}$ as the turning point for the corresponding minimal surface.

Using this fact, we obtain,
\begin{equation}
\label{xprime1}
x^{\prime}=\frac{h(z)^{-\frac{1}{2}}}{\sqrt{\left(\frac{z_t}{z}\right)^{2d-2}-1}}.
\end{equation}
Further, we introduce a dimensionless coordinate $u$ defined as $u=\frac{z}{z_{t}}$, so that the area functional $\mathcal{A}$ can be expressed as the following integral,
\begin{equation}\begin{split}
\label{area3}
\mathcal{A}=\frac{2L^{d-2}R^{d-1}}{z_{t}^{d-2}}\int_{0}^{1}\frac{du}{u^{d-1}}\frac{1}{\sqrt{h(u)\left(1-u^{2d-2}\right)}}.
\end{split}
\end{equation}

The relation between turning point $z_t$ in the bulk and the width of the strip $l$ in the boundary theory is given as,
\begin{equation}\begin{split}
\label{width}
\frac{l}{2}&=z_{t}\int_{0}^{1}\frac{u^{d-1}du}{\sqrt{h(u)\left(1-u^{2d-2}\right)}}\\
\end{split}
\end{equation}
\begin{figure}
  \centering
  \includegraphics[width=.5\linewidth]{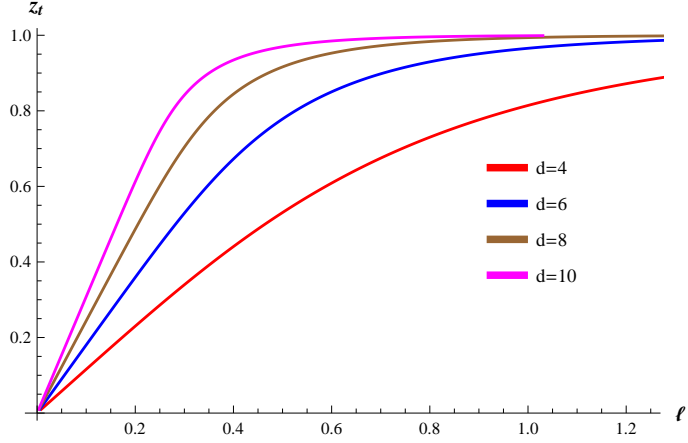}
\caption{Plot of the turning point $z_{t}$ with respect to the width $l$ for different values of the dimension $d$ given a fixed $\rho=4$ and $z_{H}=1$. This behavior indicates that the RT hypersurface extends deep inside the bulk geometry and hence gets close to the BH horizon for higher values of the dimension.}
\label{sub100}
\end{figure}
The shape of the RT surface can be best understood from the plot of the turning point $z_{t}$ as a function of the width $l$ as given in figure-1. we see that the minimal surface reaches to the horizon much quickly as a function of increasing width $l$ for higher values of dimension. Using the Euler's beta function, the above two integrations can be evaluated and the area functional takes the form as,
\begin{equation}\begin{split}\label{Afinal}
\mathcal{A}&=\frac{2}{d-2} \frac{L^{d-2} R^{d-1} }{\epsilon^{d-2}}+ \frac{2L^{d-2}R^{d-1}}{z_{t}^{d-2}}\Biggl[\frac{\Gamma{(\frac{-d+2}{2d-2})}\Gamma{(\frac{1}{2})}}{(2d-2)\Gamma{(\frac{1}{2d-2})}}
\\&+\sum_{n=1}^{\infty}\sum_{k=0}^{n}\frac{\Gamma{(n+\frac{1}{2})}\Gamma{(\frac{nd-d-k+2}{2d-2})}\rho^k(1-\rho)^{n-k}}{(2d-2)\Gamma{(k+1)}\Gamma{(n-k+1)}\Gamma{(\frac{nd-d-k+2}{2d-2}+\frac{1}{2})}}\alpha^{nd-k}\Biggr],
\end{split}
\end{equation}
where $\alpha = \frac{z_t}{z_H}$. Note that the first term in the bulk area functional given in (\ref{area3}) has some IR divergence that can be translated via holographic duality to a UV cut-off $\epsilon$ in the boundary. Similarly, the width of entangling region $l$ is obtained as,
\begin{equation}
\label{l1}
\frac{l}{2}=z_{t}\left[\sum_{n=0}^{\infty}\sum_{k=0}^{n}\frac{\Gamma{(n+\frac{1}{2})}\Gamma{(\frac{nd+d-k}{2d-2})}\rho^k(1-\rho)^{n-k}}
{(2d-2)\Gamma{(k+1)}\Gamma{(n-k+1)}\Gamma{(\frac{nd+d-k}{2d-2}+\frac{1}{2})}}\alpha^{nd-k}\right]
\end{equation}

As prescribed by the $RT$ conjecture, by using the area functional, one can compute the entanglement entropy of the strip in the boundary. To express the entanglement entropy in terms of boundary parameters by using analytical method, we need to solve  (\ref{l1}) for $z_t$ as a function of $l$. However, this procedure works only in the high and low temperature regimes. The high and low temperature regime of the boundary theory can be best understood in terms of an intrinsic length which we consider to be the width of the entangling surface $l$. By low temperature we actually mean the limit: $T_{f} \ll 1/l$ or $T_{f}l\ll1$. On the other hand the high temperature limit, with the similar line of argument can be realized by the limit $T_{f}l\gg 1$.
We have already observed that the thermal entropy of the deformed AdS blackhole is proportional to $T_{f}^{d-1}$. Therefore while defining the high and low temperature limit in obtaining the analytic expression of entanglement entropy we would expect the effective temperature $T_f$ is more appropriate as compared to the original temperature $T$. In the absence of heavy quarks, $T_{f} $ and $ T$ coincide with each other.
\subsection{EE at low effective temperature}
In the low effective temperature limit, most of the contribution for holographic entanglement entropy comes from entangling surface which has a turning point very close to the boundary. This can be effected by taking $z_{t}\ll z_{H}$ or $\alpha\ll 1$. Keeping terms up to order $\alpha^{d}$ in the low effective temperature limit, one can solve equation (\ref{l1}) for $z_{t}$ to get,
\begin{equation}\label{l2}
\begin{split}
\frac{1}{z_{t}}=\frac{2}{l}\Biggl[\left(\sqrt{\pi}\frac{\Gamma{(\frac{d}{2d-2})}}{\Gamma{(\frac{1}{2d-2})}}\right)+\rho\left(\frac{\sqrt{\pi}}{2d(2d-2)}
\frac{\Gamma{(\frac{1}{2d-2})}}{\Gamma{(\frac{d}{2d-2})}}\right)\alpha^{d-1}+(1-\rho)\left(\frac{\sqrt{\pi}}{2(d+1)}\frac{\Gamma{(\frac{2d}{2d-2})}}
{\Gamma{(\frac{1}{d-1}+\frac{1}{2})}}\right)\alpha^{d}+\mathcal{O}(\alpha^{2d-2})\Biggr]
\end{split}
\end{equation}

We solve (\ref{l2}) for $z_t$ in terms of $l$ by using a perturbative method that suggests solving the equation at any order of $\alpha$ and using the obtained result as an input at the next order.

Proceeding this way we obtain the final expression for $z_{t}$ given as,
\begin{equation}\begin{split}\label{zt}
z_{t}&=\frac{l}{2}\left(\frac{\Gamma{(\frac{1}{2d-2})}}{\sqrt{\pi}\Gamma{(\frac{d}{2d-2})}}\right)\Biggl[1-\frac{1}{2d(2d-2)}
\left(\frac{\Gamma{(\frac{1}{2d-2})}}{\Gamma{(\frac{d}{2d-2})}}\right)^{2}
\left(\frac{l}{2z_{H}\left(\sqrt{\pi}\frac{\Gamma{(\frac{d}{2d-2})}}{\Gamma{(\frac{1}{2d-2})}}\right)}\right)^{d-1}\rho \\&-\frac{1}{2(d+1)}
\left(\frac{\Gamma{(\frac{2d}{2d-2})}\Gamma{(\frac{1}{2d-2})}}{\Gamma{(\frac{d}{2d-2})}\Gamma{(\frac{1}{d-1}+\frac{1}{2})}}\right)
\left(\frac{l}{2z_{H}\left(\sqrt{\pi}\frac{\Gamma{(\frac{d}{2d-2})}}{\Gamma{(\frac{1}{2d-2})}}\right)}\right)^{d}(1-\rho)+
\mathcal{O}\left((l/z_{H})^{2d-2}\right)
\Biggr].
\end{split}
\end{equation}

The area functional in the low effective temperature up to order $(z_{H}l)^{d}$ is given as,
\begin{equation}
\mathcal{A}=A_0+R^{d-1}\left(\frac{L}{l}\right)^{d-2}S_0\Biggl(1+\rho S_2\left(\frac{l}{z_{H}}\right)^{d-1}+(1-\rho)S_1\left(\frac{l}{z_{H}}\right)^d+\mathcal{O}\left((l/z_{H})^{2d-2}\right)\Biggr),
\end{equation}
where, $A_{0}$ is the diverging term and the other constant terms $S_0$, $S_1$, $S_2$ are given below,
\begin{equation}\begin{split}\label{Snotonetwo}
S_0&=\frac{2^{d-2}\pi^{\frac{d-1}{2}}\Gamma{\left(\frac{-d+2}{2d-2}\right)}}{(d-1)\Gamma{\left(\frac{1}{2d-2}\right)}}\left(\frac{\Gamma{\left(\frac{d}{2d-2}\right)}}
{\Gamma{\left(\frac{1}{2d-2}\right)}}\right)^{d-2}\\
S_1&=2^{-(d+1)}\pi^{-\frac{d}{2}}\frac{\Gamma{\left(\frac{1}{2d-2}\right)^{d+1}}}{\Gamma{\left(\frac{1}{d-1}+\frac{1}{2}\right)}\Gamma{\left(\frac{d}{2d-2}\right)^d}}
\left[\frac{\Gamma{\left(\frac{1}{d-1}\right)}}{\Gamma{\left(\frac{-d+2}{2d-2}\right)}}+\frac{(d-2)}{(d+1)}\frac{2^{\frac{1}{d-1}}}
{\sqrt{\pi}}\Gamma{\left(1+\frac{1}{2d-2}\right)}\right]\\
S_2&=2^{-d}\pi^{-(\frac{d-1}{2})}\left(\frac{\Gamma{\left(\frac{1}{2d-2}\right)}}{\Gamma{\left(\frac{d}{2d-2}\right)}}\right)^{d+1}\left[
\frac{\Gamma{\left(\frac{d}{2d-2}\right)}}
{\Gamma{\left(\frac{-d+2}{2d-2}\right)}}+\frac{(d-2)}{2d(d-1)}\right]
\end{split}
\end{equation}

Using the RT proposal, we finally express the holographic entanglement entropy of the boundary strip within the low- effective temperature regime.
\begin{equation}
\label{lowS}
\mathcal{S}_{\mathcal{A}}=\frac{2}{d-2} \frac{L^{d-2} R^{d-1} }{\epsilon^{d-2}}+\frac{R^{d-1}}{4G^{d+1}_{N}}\left(\frac{L}{l}\right)^{d-2}S_0\Biggl(1+\rho S_2\left(\frac{4\pi T_{f}l}{d}\right)^{d-1}+(1-\rho)S_1\left(\frac{4\pi T_{f}l}{d}\right)^d+\mathcal{O}\left((T_{f}l)^{2d-2}\right)\Biggr).
\end{equation}
We have a few remarks regarding the final form of the holographic entanglement entropy at low temperature. In order to analyze the result obtained for holographic entanglement entropy at low temperature limit we first enunciate the fact that $ \rho = d (1- \frac{T}{T_f})$ signifies a dimensionless quantity that scales as $\mathcal{O}(1)$. Hence it does not contribute to any power of $T_f$ in all correction terms present in the expression for holographic entanglement entropy.
\begin{itemize}
\item Note that, the final expression of holographic entanglement entropy in the low temperature regime is modified due to the presence of heavy quark as compared to the one obtained for strongly coupled SYM plasma at finite temperature \cite{Fischler:2012ca}. A similar modification is also observed in \cite{Kundu:2016dyk} for charged strongly coupled plasma where the leading order correction is proportional to the $d$th power of some effective temperature defined appropriately for charged plasma. However, in our case, the leading order correction is proportional to $T_f^{d-1}$. The reason for this difference can be understood from the behavior of the blackening function $h(z)$ for the deformed AdS-blackhole background given in \ref{h}. The extremal surface relevant for the holographic computation of EE at low effective temperature lies close to the boundary where the leading term in the blackening function behaves as $z^{d-1}$ and hence dominates over the mass term proportional to $z^d$. Hence it is expected that in the low effective temperature limit, the maximum contribution to EE comes from terms involving quark density, which turns out to be proportional to $ T_f^{d-1}$.
\item It is also important to note that the parameter $\rho$ we have introduced here varies from $0$ to $d$ and also it is proportional to the quark density $b$. Now to retain the appropriate zero quark density ($\rho \rightarrow 0$) limit in  EE, it is necessary to keep the sub-leading contribution proportional to $T_f^{d}$ in the expression of EE.

\end{itemize}

\subsection{EE at high effective temperature limit}
The high temperature limit of holographic entanglement entropy can be realized by choosing the RT surface approaching near the horizon of the bulk geometry. In other words, in this case we must consider the limit $z_{t}\rightarrow z_{H}$ in order to see the high temperature effects. However, in this limit, the series expansions of the integrands in \ref{area3} and in \ref{width} we have explicitly used to obtain both $l$ and $\mathcal{A}$, become divergent. Nevertheless,  the following combination of $\mathcal{A}$ and $l$ is finite and well-behaved.

\begin{equation}
\label{highA}
\begin{split}
\mathcal{A}-\frac{L^{d-2}R^{d-1}}{z_{t}^{d-1}}l=
\frac{2L^{d-2}R^{d-1}}{z_{t}^{d-2}}\int_{0}^{1}du\frac{\sqrt{h(u)}}{\sqrt{1-u^{2d-2}}}\frac{(1-u^{2d-2})}{u^{d-1}}
\end{split}
\end{equation}

Using the prescribed above combination  \ref{highA}, we express the finite part of the area functional,
\begin{equation}
\begin{split}
\mathcal{A}&=\frac{L^{d-2}R^{d-1}}{z_{t}^{d-1}}l\\&+\frac{2L^{d-2}R^{d-1}}{z_{t}^{d-2}}\left\{\frac{\sqrt{\pi}
\Gamma{(-\frac{d-2}{2d-2})}}{(2d-2)\Gamma{(\frac{1}{2d-2})}}+\int_{0}^{1}du\left(\frac{\sqrt{1-u^{2d-2}}}{u^{d-1}\sqrt{h(u)}}
-\frac{1}{u^{d-1}
\sqrt{1-u^{2d-2}}}\right)\right\},
\end{split}
\end{equation}
which is finite in the limit $z_{t}\rightarrow z_{H}$ as anticipated. Replacing $z_{H}$ by the temperature $T_{f}$ we get the final result for the EE at high temperature as,
\begin{equation}
\begin{split}
\label{highS}
\mathcal{S}_{\mathcal{A}}\approx \frac{R^{d-1}}{4G^{d+1}_{N}}\Biggl[
V\left(\frac{4\pi T_{f}}{d}\right)^{d-1}\Biggl\{1+2\left(\frac{d}{4\pi T_{f} l}\right)\widetilde{S}(\rho,d)\Biggr\}\Biggr],
\end{split}
\end{equation}
where $V=l L^{d-2}$ is the $(d-1)$ dimensional volume of the strip and $\widetilde{S}(\rho,d)$ is given as,
\begin{equation}
\begin{split}\label{stilde}
\widetilde{S}(\rho,d)=\Biggl\{\frac{\sqrt{\pi}
\Gamma{(-\frac{d-2}{2d-2})}}{(2d-2)\Gamma{(\frac{1}{2d-2})}}+\int_{0}^{1}du\left(\frac{\sqrt{1-u^{2d-2}}}{\sqrt{h(u)}u^{d-1}}-\frac{1}{u^{d-1}
\sqrt{1-u^{2d-2}}}\right)\Biggr\}.
\end{split}
\end{equation}
\begin{figure}
  \centering
  \includegraphics[width=.5\linewidth]{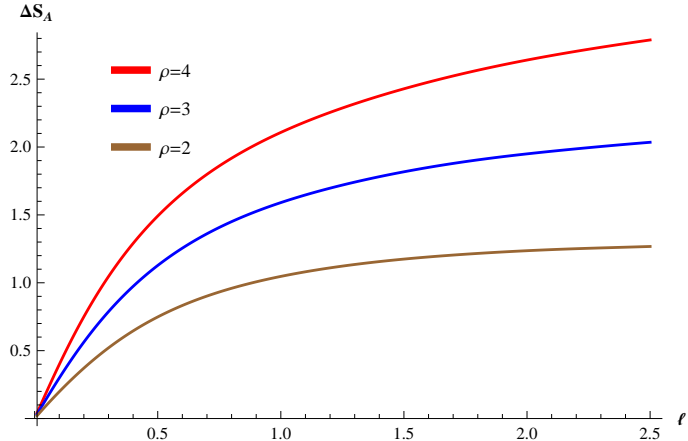}
  \caption{Plot of $\Delta S_{A}=\left(S_{A}(T_{f},\rho)-S_{A}(T_{f}=0,\rho=0)\right)$ as a function of width $l$ for different values of $\rho$. The plot clearly shows increasing behavior of EE at finite temperature as $\rho$ increases. }
 \label{sub45}
\end{figure}
\begin{itemize}
\item Note that EE at high effective temperature we obtain in (\ref{highS}) contains leading order term proportional to $T_{f}^{d-1}$ which is exactly similar to the behavior of thermal entropy as computed in section-${2}$. Since at high temperature, the boundary field theory is in the thermal regime, the maximum contribution to the EE comes from the thermal fluctuations. From the perspective of dual bulk gravity, taking the high temperature limit implies that the RT surface extends to the region which is very close to the horizon. Hence the similarity between the thermal entropy and the EE in the high effective temperature limit is expected and also agree with the result obtained in  \cite{Kundu:2016dyk}.
\item We observe that in high temperature regime, the leading order correction in the finite part of the holographic entropy is independent of heavy quark density $b$. However, the sub leading contribution depends on $b$ via the parameter $\rho$. We present this sub leading correction to the finite part of HEE for $d = 4$.
\begin{equation}\begin{split}\label{rhodiff}
\widetilde{S}(\rho=0)=-0.33,~~ \widetilde{S}(\rho=1)=-0.024,~~\widetilde{S}(\rho=2)=0.32,~~\widetilde{S}(\rho=3)=0.75,~~\widetilde{S}(\rho=4)=1.67.\nonumber
\end{split}
\end{equation}
For a specific choice of dimension $d=4$, the dimensionless parameter $\rho$ solely depends on the density of heavy quarks. Now, as we increase the value of quark density, $\rho$ also increases. Correspondingly, the sub leading contribution to the finite part of EE also monotonically increases. We noted similar behavior of $\widetilde{S}$ for $d=5$ and $d=6$.

\end{itemize}
\section{Entanglement wedge cross section in $d$ dimension}
Generally, a more appropriate microscopic description of a strongly coupled large $N_c$ gauge theory at finite temperature is given by mixed entangled state which carries the information of both classical and quantum correlation.
RT conjecture for black hole background estimates the both thermal and quantum correlation in the dual strongly coupled large $N_c$ gauge theory at finite temperature and within specific approximation the final result clearly attributes to the entanglement entropy as well as the thermal entropy.  However, extracting the sole contribution of the quantum correlation for the mixed bipartite entanglement for a strongly coupled field theory has been awaited for long time. Recently a new conjecture is proposed to study entanglement of purification as a suitable measure of mixed entanglement of a strongly coupled field theory by virtue of a computing a novel holographic dual called entanglement wedge cross section \cite{Takayanagi:2017knl, Nguyen:2017yqw}.

\begin{figure}
\centering
 \includegraphics[width=.75\linewidth]{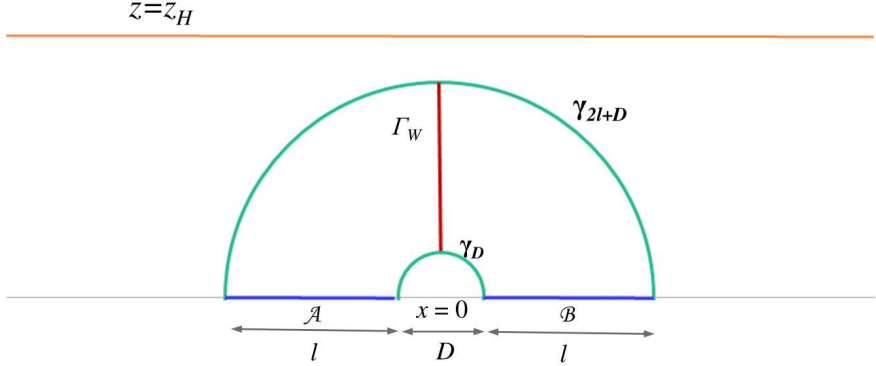}
  \caption{Schematic diagram of two disjoint subregions of width $l$ separated by a distance $D$ for
the computation of EWCS.}
 \label{sub111}
\end{figure}

To define the entanglement wedge, one needs two non overlapping subsystems $\mathcal{A}$ and $\mathcal{B}$ on the boundary of some bulk geometry $\mathcal{M}$.  Let's denote the minimal RT surface for the region $(\mathcal{A}\cup \mathcal{B})$ by $\gamma_{\mathcal{A}\mathcal{B}}$ such that $\gamma_{\mathcal{A}\mathcal{B}}\equiv \left(\gamma_{2l+D}\cup \gamma_{D}\right)$ (see figure-\ref{sub111} for details). Then the entanglement wedge is defined by the volume of the bulk geometry with boundary $(\mathcal{A}\cup \mathcal{B}\cup\gamma_{\mathcal{A}\mathcal{B}})$. The entanglement wedge cross section is the minimal area surface $\Gamma_{W}$ that completely separates the two subregions $ \mathcal{A}$, $ \mathcal{B}$ with it's boundary ending on $\gamma_{\mathcal{A}\mathcal{B}}$. In the following, we calculate the entanglement wedge cross section in a $d+1$ dimensional deformed AdS-BH  geometry dual to $d$ dimensional strongly coupled large $N_c$ gauge theory at finite temperature in the presence of a uniform heavy quark density by using the holographic prescription presented in \cite{BabaeiVelni:2019pkw, Jokela:2019ebz}.
Here we consider two parallel infinitely long strips with equal width $l$ describing the two subregions $\mathcal{A}$ and $\mathcal{B}$ which are separated by a distance $D$. The two subregions are considered in a particular configuration which is symmetric around $x=0$, $x$ being one of the spatial directions such that,
\begin{equation}
\begin{split}
\mathcal{A}&=\left\{l+D/2>x>D/2; -L/2<y_{i}<L/2, i=2,3,...., d-2 \right\}\\
\mathcal{B}&=\left\{-l-D/2<x<-D/2; -L/2<y_{i}<L/2, i=2,3,...., d-2\right\}.
\end{split}
\end{equation}
In this configuration the minimal area surface $\Sigma_{min}$ that separates $\mathcal{A}$ and $\mathcal{B}$ will be given by the vertical surface at $x=0$ (see \cite{Liu:2019qje} for asymmetric choice of cross section). The induced metric on this constant time slice is given as,
\begin{equation}
ds_{\Sigma_{min}}^2=\frac{R^2}{z^2}\left(d\vec{x}_{d-2}^2+\frac{dz^2}{h(z)}\right),
\end{equation}
Then the entanglement wedge cross section can be calculated as,
\begin{equation}\begin{split}\label{EW}
E_{W}&=\frac{L^{d-2}R^{d-1}}{4G^{d+1}_{N}}\int^{z_{t}(2l+D)}_{z_{t}(D)}\frac{dz}{z^{d-1}\sqrt{h(z)}}\\
&=\frac{L^{d-2}R^{d-1}}{4G^{d+1}_{N}}\sum_{n=0}^{\infty}\sum_{k=0}^{n}\left(\frac{1}{nd-d-k+2}\right)\frac{\Gamma{(n+\frac{1}{2})}\rho^{k}(1-\rho)^{n-k}}
{\Gamma{(k+1)}\Gamma{(n-k+1)}\Gamma{(\frac{1}{2})}}\\&
\times\Biggl\{\frac{z_{t}(2l+D)^{nd-d-k+2}}{z_{H}^{nd-k}}-\frac{z_{t}(D)^{nd-d-k+2}}{z_{H}^{nd-k}}\Biggr\}.
\end{split}
\end{equation}
In the previous section while studying the low and high effective temperature behaviour of EE, we had only one length scale corresponding to the width of the rectangular strip-like entangling region and the two limiting temperatures are defined whether $T_{f}\ll\frac{1}{l}$ or $T_{f}\gg\frac{1}{l}$. However, in this computation there is another length scale $D$ corresponding to the separation between the two subregions as previously mentioned, and hence the following set of choices can be considered: (i) $T_{f}\ll \frac{1}{l},~T_{f}\ll \frac{1}{D}$, which is the usual low effective temperature limit where the temperature is much smaller in comparison to both the length scales associated to $l$ and $D$, (ii) $T_{f}\gg \frac{1}{l},~T_{f}\ll\frac{1}{D}$ under which the effective temperature is large in comparison to the length scale associated to $l$ but it is small with respect to the length scale associated to $D$. So this limit corresponds to the usual high temperature limit. Lastly, we mention the third possibility defined as (iii) $DT_{f}\gg lT_{f}$ or $T_{f}\gg\frac{1}{l},~T_{f}\gg\frac{1}{D}$. This particular limit is not relevant for our purpose since it corresponds to a disentangling phase to two subregions and the EWCS becomes identically equal to zero and thus we exclude it from our present analysis.
\subsection{EWCS in low effective temperature}
As explained above, to obtain the analytic form of EWCS in low effective temperature limit we follow $D T_{f}\ll lT_{f}\ll 1$. In this case the turning points for both of the RT surfaces, $\gamma_{D}$ and $\gamma_{2l+D}$ lies far away from the horizon of the black hole, $z_{t}\ll z_{H}$. Hence
one can ignore all the higher order terms and terminate the infinite series in (\ref{EW}) up to the order $\left(1/z_{H}\right)^{d}$. Now using the approximate expression for both $z_{t}(D)$ and $z_{t}(2l+D)$ as given in (\ref{zt}) the final result for EWCS is given as,
\begin{equation}
\begin{split}\label{lowEW}
{E_{W}}^{\text{low}}=\frac{L^{d-2} R^{d-1}}{4G^{d+1}_{N}} \Biggl\{\mathcal{E}_{0}
   \left(\frac{1}{D^{d-2}}-\frac{1}{(D+2 l)^{d-2}}\right)+\rho\mathcal{E}_{1} l \left(\frac{4\pi T_{f}}{d}\right)^{d-1}-(1-\rho )\mathcal{E}_{2}l(l+D)\left(\frac{4\pi T_{f}}{d}\right)^{d}
   \Biggr\},
\end{split}
\end{equation}
where $\mathcal{E}_{0}$, $ \mathcal{E}_{1}$, $\mathcal{E}_{2}$ are all $\mathcal{O}(1)$ constants and only depends on $d$ with the following explicit expressions,
\begin{equation}
\label{lowEw}
\begin{split}
\mathcal{E}_{0}&=\frac{2^{d-2} \pi ^{\frac{d-2}{2}}}{d-2} \left(\frac{\Gamma
   \left(\frac{d}{2 d-2}\right)}{\Gamma
   \left(\frac{1}{2 d-2}\right)}\right)^{d-2}\\
   \mathcal{E}_{1}&=\frac{1}{2\sqrt{\pi}}\left(\frac{\Gamma \left(\frac{1}{2 d-2}\right)}{\Gamma
   \left(\frac{d}{2 d-2}\right)}-\frac{\Gamma
   \left(\frac{1}{2 d-2}\right)^3}{d (2 d-2) \Gamma
   \left(\frac{d}{2 d-2}\right)^3}\right)\\
   \mathcal{E}_{2}&=\left(\frac{1}{2\pi(d+1)}\left(\frac{\Gamma
   \left(\frac{1}{2 d-2}\right)}{\Gamma
   \left(\frac{d}{2 d-2}\right)}\right)^3 \frac{\Gamma
   \left(\frac{d}{d-1}\right)}{\Gamma
   \left(\frac{d+1}{2d-2}\right)}-\frac{\Gamma \left(\frac{1}{2
   d-2}\right)^2}{4 \pi  \Gamma \left(\frac{d}{2
   d-2}\right)^2}\right).
   \end{split}
\end{equation}
As expected, the first term in the right hand side of (\ref{lowEw}) increases as the separation distance $D$ between the two subregions decreases and in the limit $D\rightarrow 0$, EWCS diverges. In the following, we make some comments regarding the correction terms in the analytical expression of ${E_{W}}^{\text{low}}$ due to the back reaction on the AdS BH geometry.
\begin{itemize}
\item We observe the leading order correction term appearing in (\ref{lowEw}) is proportional to the dimensionless quark density parameter $\rho$. Similar to the result of entanglement entropy at low effective temperature, this leading order correction term is proportional to the volume of the entangling hypersurface. Also at a fixed temperature the above result for EWCS shows increasing behavior with $\rho$.
    \item However, unlike the behavior of entanglement entropy, in the limit of vanishing quark density, the EWCS decreases with increasing effective temperature $T_{f}$. Also note that in the same vanishing quark density limit, the above result at low effective temperature correctly reproduces the result obtained earlier for the AdS-BH geometry in \cite{BabaeiVelni:2019pkw}.
    \item Again as already observed in the EE calculation, to make the zero quark density ($\rho \rightarrow 0$) limit appropriate, we also require to keep a sub-leading correction term which is proportional to $T_f^{d}$.
\end{itemize}
\subsection{EWCS in high effective temperature}
As previously mentioned, the sensible way to take the high temperature limit along with maintaining the entanglement between $\mathcal{A}$ and $\mathcal{B}$ follows the inequality $DT_{f}\ll 1\ll lT_{f}$ which translates to considering the following two approximations: (i) $z_{t}(D)\ll z_{H}$ and (ii) $z_{t}(2l+D)\rightarrow z_{H}$. Hence using the first approximation among the above two, we can replace the infinite series in (\ref{EW}) associated to the second term inside the curly bracket by the leading order term in $\left(1/z_{H}\right)$ as effected by considering $n,k=0$. However for infinite series related to the first term inside the curly bracket, we must ensure it's convergence for large values of $n$ as $z_{t}(2l+D)$ approaches $z_{H}$. To do this we first evaluate the sum over the index $k$ and then consider the low $\rho$ limit to get the infinite sum as,
\begin{figure}

  \centering
  \includegraphics[width=.5\linewidth]{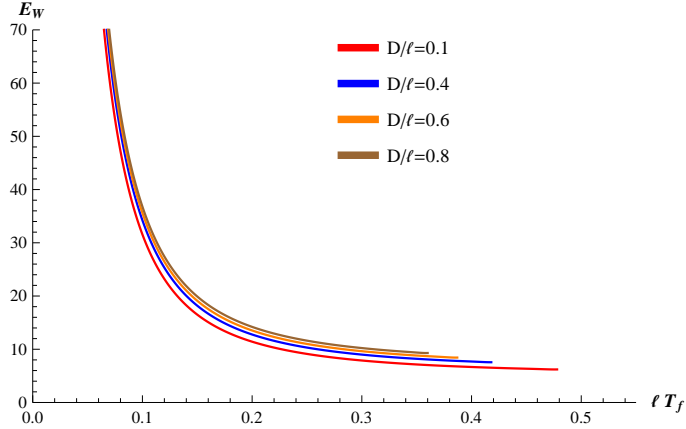}
  \caption{Variation of $E_{W}$ with respect to the dimensionless quantity $lT_{f}$ for different values of $D/l$, indicating the decreasing behavior of EWCS as the effective temperature increases (with $l=1$).}
 \label{fig:1}

\end{figure}
\begin{equation}
\begin{split}
\sum_{n=0}^{\infty}\left\{\frac{\Gamma \left(n+\frac{1}{2}\right)}{(nd-d+2)\Gamma (n+1)}(1-n\rho)+\frac{(2 n+1) \Gamma \left(n+\frac{1}{2}\right)}{2 (d n+1)
   \Gamma (n+1)}\left(\frac{z_{H}}{z_{t}}\right)\rho+\mathcal{O}(\rho^2)\right\}\frac{z_{t}(2l+D)^{nd-d+2}}{z_{H}^{nd}}.
   \end{split}
\end{equation}
Now in the large $n$ limit the above infinite sum behaves as,
\begin{equation}
\begin{split}\label{ee}
\frac{\rho}{d}\frac{1}{\sqrt{n}}\frac{z_{t}(2l+D)^{nd-d+1}}{z_{H}^{nd-1}}-\frac{\rho}{d}\frac{1}{\sqrt{n}}\frac{z_{t}(2l+D)^{nd-d+2}}{z_{H}^{nd}}
+\mathcal{O}\left(\frac{1}{n^{3/2}}\frac{z_{t}(2l+D)^{nd-d+2}}{z_{H}^{nd}}\right).
\end{split}
\end{equation}
We notice that the first two terms individually are not convergent as each of them varies as $1/\sqrt{n}$. However due to the presence of the relative sign between the first two terms in (\ref{ee}), such divergences eventually get canceled to give finite result in the large $n$ limit.

Now to test the finiteness of $E_{W}$ for arbitrary $\rho$, we consider a particular dimension $d=4$ and hence the maximum value allowed for $\rho$, becomes four. For this chosen value of $d$, it is possible to exactly evaluate the integral in (\ref{EW}). We have plotted the result as a function of the dimensionless variable $l T_{f}$ for four different values of $\frac{D}{l}$ (see Figure \ref{fig:1}). In this plot we observe that for each value of the ratio $\frac{D}{l}$, the wedge cross section converges to a finite value. We also see that the smaller the ratio $\frac{D}{l}$ is, the lower the corresponding cross-section becomes. Notice that as the ratio $\frac{D}{l}$ decreases and approaches to zero, the condition for high effective temperature limit mentioned above is attained more accurately. Since EWCS is a monotonically increasing function of $\rho$ and it turns out to be finite for the maximum allowed value of $\rho$ for a given particular dimension $d=4$, we conclude that for arbitrary values of $\rho$, the wedge cross section in high temperature limit must converge to a finite value. The final result of the entanglement wedge cross section at high temperature is given as,
\begin{equation}
\begin{split}\label{highEW}
{E_{W}}^{high}&=\frac{L^{d-2} R^{d-1}}{4G^{d+1}_{N}}T_{f}^{d-2}\Biggl\{-
   \left(\frac{2\sqrt{\pi}\Gamma \left(\frac{d}{2
   d-2}\right)}{\Gamma \left(\frac{1}{2
   d-2}\right)}\right)^{d-2}\left(\frac{1}{DT_{f}}\right)^{d-2}+\mathcal{C}\left(\frac{4\pi}{d}\right)^{d-2}\\&- \rho \left(\frac{d-2}{8 \sqrt{\pi } d(d-1) }\right)\left(\frac{\Gamma \left(\frac{1}{2 d-2}\right)}{\Gamma
  \left(\frac{d}{2 d-2}\right)}\right)^{3}  \left(\frac{4 \pi }{d}\right)^{d-1}\left(DT_{f}\right)\\&-(1-\rho)\left(\frac{d-2}{8\pi(d+1)}\right)\frac{\Gamma
\left(\frac{2 d}{2 d-2}\right)}{\Gamma
\left(\frac{1}{2}+\frac{1}{d-1}\right)}\left(\frac{\Gamma \left(\frac{1}{2 d-2}\right)}{ \Gamma
\left(\frac{d}{2 d-2}\right)}\right)^{3} \left(\frac{4 \pi}{d}\right)^{d} \left(DT_{f}\right)^2 \Biggr\},
\end{split}
\end{equation}
where we introduce $\mathcal{C}$ as,
\begin{equation}\
\mathcal{C}=\sum_{n=0}^{\infty}\sum_{k=0}^{n}\left(\frac{1}{nd-d-k+2}\right)\frac{\Gamma{(n+\frac{1}{2})}\rho^{k}(1-\rho)^{n-k}}
{\Gamma{(k+1)}\Gamma{(n-k+1)}\Gamma{(\frac{1}{2})}}
\end{equation}
\begin{figure}
\centering

  \centering
  \includegraphics[width=.5\linewidth]{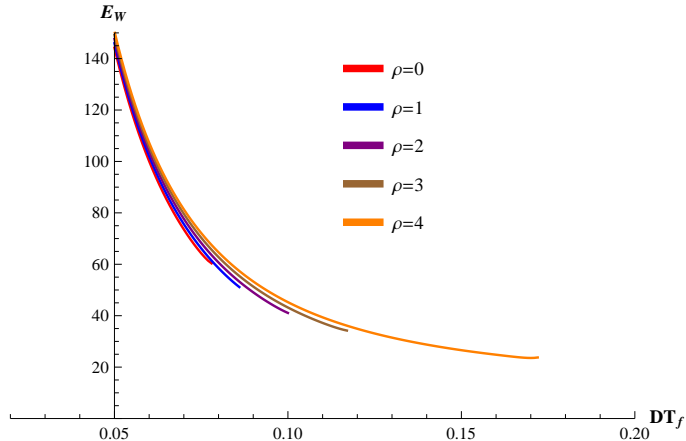}
  \caption{Variation of $E_{W}$ with dimensionless quantity $DT_{f}$ for different values of $\rho$. Again we see the increasing behavior of EWCS with $\rho$ which is similar to the variation of EE.}

\label{fig4}
\end{figure}
The behavior of the ${E_{W}}^{high}$ with respect to the presence of back reaction in the bulk geometry is summarized as:
\begin{itemize}
\item The most significant term in the above result (\ref{highEW}) is the second term that varies as the area of the subregions even at finite temperature. Contrary to the results obtained for the holographic entanglement entropy at high effective temperature (\ref{highS}) which accommodates a volume dependance for the leading order correction term, here we observe EWCS contains an area dependent leading order correction term. Similar observation is perviously reported in \cite{BabaeiVelni:2019pkw}.
\item It is important to note that at both high and low temperature limits, EWCS increases with the quark density $\rho$ which can be explicitly shown by plotting the measure as a function of dimensionless quantity $DT_{f}$ for different $\rho$ (see fig-4). Also from the same plot it is clear that the critical distance of separation between the two subregions increases with $\rho$ (we will see this again in the next subsection), thus corresponds to the increase of entanglement and hence the EWCS.
    \item Further, for both regime of the effective temperature, the variation of  EWCS  with $lT_{f}$ for different values of $D/l$ as shown in figure \ref{fig:1}, indicates that the temperature at which the quantity drops discontinuously to zero increases as the ratio $D/l$ decreases.

\end{itemize}
\subsection{Critical distance between the strips}
In this section we will study the phase transition occurring between two spatially disjoint subregions $\mathcal{A}$ and $\mathcal{B}$. The transition we are going study here happens from a bipartite entangled phase associated to the direct product of the Hilbert spaces of region $\mathcal{A}$ and $\mathcal{B}$ respectively to the one with zero entanglement due to the increase of separation between $\mathcal{A}$ and $\mathcal{B}$ beyond a particular critical value.

For pure AdS geometry this kind of phase transition occurs as the ratio $D/l$ becomes greater than a critical value. However for AdS-BH background, if the distance of separation $D$ is greater than a particular critical value $D_c$ then the two subregions will always be in a disentangled phase irrespective of the width $l$. One interesting aspect of EWCS is that it can capture this phase transition such that in a disentangled phase, the EWCS drops discontinuously to zero value \cite{BabaeiVelni:2019pkw, Jokela:2019ebz}.

In this regard we emphasize that as described in figure \ref{fig4}, for each choice of $\rho$, EWCS behaves as a monotonically decreasing function of D(assuming $T_f$ = 1) and it discontinuously drops down to zero value until $D$ reaches to a critical value $D_c$. Moreover, from the same figure we note that $D_c$ is a monotonically increasing function of $\rho$.  Alternatively, similar behavior of $D_c$ with respect to $\rho$ can be studied by analyzing the holographic mutual information(HMI) as a measure of the total correlation of the theory.
It turns out that as $D$ approaches to the critical value, the corresponding HMI decreases and finally becomes zero, however in a continuous manner unlike the behavior of EWCS at around $D_c$. HMI of the two subregions $\mathcal{A}$ and $\mathcal{B}$ is given as,
\begin{eqnarray}
&&I (l, D) = 2S(l)-S(2l+D)-S(D). \nonumber \\
&&I (l, D_c) = 0
\end{eqnarray}
\begin{figure}
\centering
\begin{subfigure}{.45\textwidth}
  \centering
  \includegraphics[width=1\linewidth]{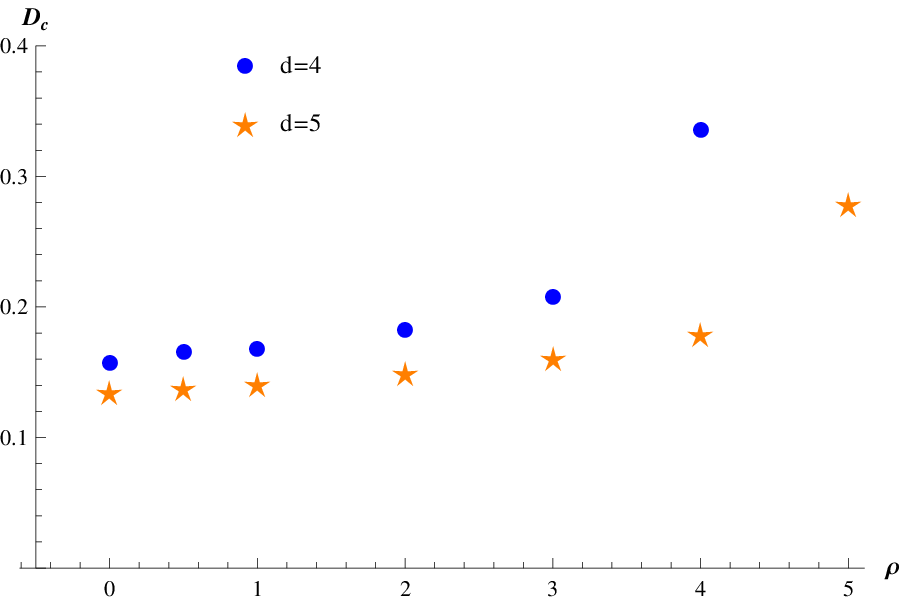}
\caption{}
 \label{sub1}
\end{subfigure}
\begin{subfigure}{.45\textwidth}
 \includegraphics[width=1\linewidth]{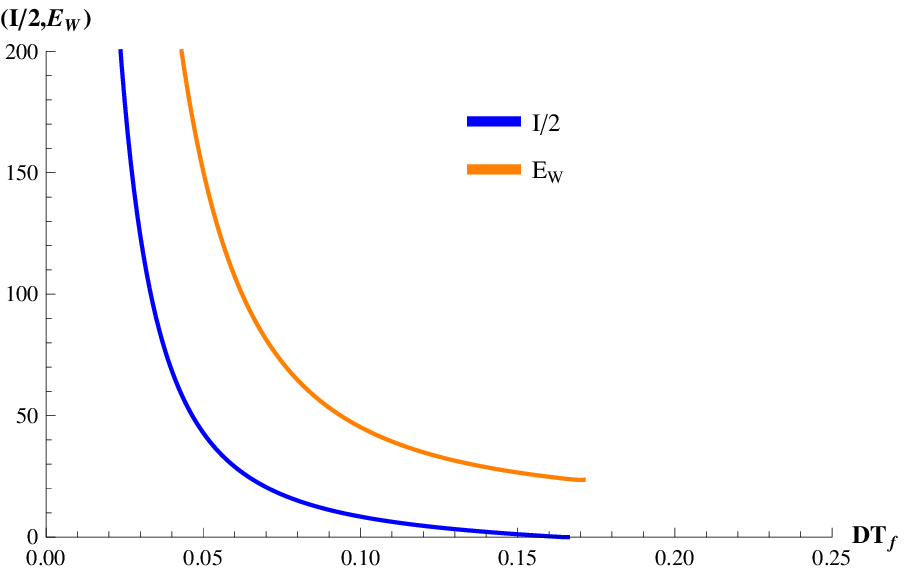}
  \caption{}
  \label{sub2}
\end{subfigure}
\caption{(A) The critical distance $D_{c}$ between two specially disjoint regions increases with $\rho$ for two values of the dimension $d=4,5$. (B) The corresponding plot shows the variation of $E_{W},I/2$ with temperature for a fixed $\rho=4$ and dimension $d=4$. Both $E_{W}$ and $I$ show decreasing behavior with temperature.}
\end{figure}
Since this phase transition phenomenon happens for a large value of strip width $l$, one can approximate $S(l)$ and $S(2l+D)$ by the earlier obtained analytic expressions at high effective temperature (\ref{highS}), while for $S(D)$ we will use the corresponding low effective temperature result (\ref{lowS}), to get the required condition as,
\begin{equation}
\label{critical}
\left(2 \frac{\widetilde{S}(d, \rho)}{z_ {H}^{d - 2}}-\frac{D_{c}}{z_{H}^{d - 1}}\right)-\frac{S_{0}}{D_{c}^{d - 1}}-\rho
S_0 S_2\frac{D_{c}}{z_{H}^{d-1}}-\left(1 -\rho\right)
S_{0} S_{1}\frac {D_{c}^{2}} {z_{H}^{d}}=0,
\end{equation}
where for explicit expression of $\widetilde{S}$, $S_{0}$, $S_{1}$ and $S_{2}$ (see \ref{stilde}, \ref{Snotonetwo} respectively). In order to obtain the critical value of distance of separation $D_c$ we solve the above equation for two different choices of dimension, i.e $d=4$ and $d=5$. In fig ({\ref{sub1}}), we have plotted the critical distance $D_c$ as a monotonically increasing function of $\rho$ for both $d=4$ and $d=5$. As we can see from the plot that the outcome of this alternative analysis associated to HMI reconfirms that $D_c$ increases monotonically with respect to $\rho$. Finally, to verify the validity of the inequality between HMI and the EWCS as mentioned in (\ref{MIEOP}), we have plotted both $I/2$ and $E_{W}$ as a function of a dimensionless temperature $T_{f}D$ in fig ({\ref{sub2}}). It is clearly evident from the above figure that the plot which corresponds to the variation of $I/2$ as a function of $T_{f}D$ (in blue) lies below the plot showing the variation of EWCS (in brown) for every possible values of the dimensionless parameter, thereby satisfying the inequality in (\ref{MIEOP}). Both HMI and EWCS decrease as $T_{f}D$ increases and they approach to zero beyond a particular value of $D$($T_f = 1$) indicating the transition point as already mentioned.

\section{Subregion volume complexity at finite quark density}
Ryu-Takayanagi prescription provides us a nice holographic prescription to study the entanglement structure of the boundary theory by analyzing the area of a co-dimension two minimal surface as a geometric measure in the dual bulk theory. It came out as a natural curiosity whether we can generalize further such connection between a bulk  geometrical quantity other than the area of RT surface and an appropriate measure associated to the quantum information in the dual boundary theory. Along this line of thought, the author in \cite{Alishahiha:2015rta} introduces the idea of considering a co-dimension one spacelike volume bounded by the co-dimension two Ryu-Takayanagi hypersurface and comes up with a proposal that the mentioned volume is holographically dual to the quantum complexity of the boundary theory. The subregion volume complexity is a particular version of the ``complexity equals volume" conjecture (previously mentioned in the introduction) and it computes the complexity of a mixed quantum state defined on the subregion (entangling region) of the boundary theory.
In this section following \cite{Alishahiha:2015rta}, we wish to compute the modification to the volume complexity of an infinite strip like subregion in the boundary theory due to the presence of the uniform distribution of heavy quarks. More precisely, the correspondence between co-dimension one space like volume in the $d+1$ dimensional bulk and the subregion volume complexity in the $d$ dimensional boundary theory is encoded as follows,
\begin{equation}
\label{complexity formula}
\mathcal{C}_{V}=\Biggl[\frac{V}{G_{N}^{(d+1)} R}\Biggr],
\end{equation}
where $R$ is the radius of the AdS spacetime and $G_{N}^{(d+1)}$ is the $d+1$ dimensional Newton's constant.
Again we define the embedding of the minimal RT surface as $x(z)$ and the corresponding enclosed volume is given as,
\begin{equation}
\label{vol}
V=2\int_{-\frac{L}{2}}^{\frac{L}{2}} d^{d-2} y\int_{0}^{z_{t}} dz \sqrt{g}\int_{0}^{x(z)}dx=2L^{d-2} \int_{0}^{z_{t}} dz \sqrt{g} ~x(z).
\end{equation}
In the above expression, $\sqrt{g}$ is the $d$-dimensional volume element evaluated from the metric (\ref{metric3}) at constant time slice and $x(z)$ can be obtained from equation (\ref{xprime1}) as,
\begin{equation}
\label{xprime222}
x(z)=\int_{z}^{z_{t}} dz^{\prime}\frac{h(z^{\prime})^{-\frac{1}{2}}}{\sqrt{\left(\frac{z_t}{z^{\prime}}\right)^{2d-2}-1}}.
\end{equation}
In terms of the dimensionless variable $u$ defined in the previous sections, the above integral can be re-casted as,
\begin{equation}
V=\frac{2L^{d-2}R^{d}}{z_{t}^{d-2}}\int_{0}^{1}\frac{1}{\sqrt{h(u)}u^d}du\int_{u}^{1}\frac{u'^{d-1}}{\sqrt{h(u')(1-u'^{2d-2})}}du',
\end{equation}
One can split the second integral in the RHS and use the integral form of $l$ as given in (\ref{area3}) to rewrite the above as,
\begin{equation}
V=\frac{2L^{d-2}R^{d}}{z_{t}^{d-2}}\Biggl[\underbrace{\frac{l}{2z_{t}}\int_{0}^{1}\frac{1}{\sqrt{h(u)}u^d}du}_{I_{1}}-
\underbrace{\int_{0}^{1}\left(\frac{1}{\sqrt{h(u)}u^d}\int_{0}^{u}\frac{u'^{d-1}}{\sqrt{h(u')(1-u'^{2d-2})}}du'\right)du}_{I_{2}}\Biggr].
\label{complex}
\end{equation}
Using the series expansion of $1/\sqrt{h(u)}$ one can evaluate the above two definite integral and they are given as, 
\begin{equation}
\begin{split}\label{I11}
I_{1}&=\frac{l}{2}\left(\frac{z_{t}^{d-2}}{d-1}\frac{1}{\epsilon^{d-1}}+\frac{\rho}{2}\frac{z_{t}^{d-2}}{z_{H}^{d-1}}\log{\left(\frac{\epsilon}{z_{t}}\right)}\right)
\\&+\frac{l}{2z_{t}}\Biggl\{-\frac{1}{d-1}+\frac{1-\rho}{2}\alpha^{d}
+\sum_{n=2}^{\infty}\sum_{k=0}^{n}\frac{\Gamma{(n+\frac{1}{2})}\rho^k(1-\rho)^{n-k}\alpha^{nd-k}}{\Gamma{(k+1)}
\Gamma{(n-k+1)}\Gamma{(\frac{1}{2})}(nd-d-k+1)}\Biggr\},
\end{split}
\end{equation}

\begin{equation}
\begin{split}\label{I22}
I_{2}&=\frac{l}{2z_{t}}\sum_{p=0}^{\infty}\sum_{q=0}^{p}\left(\frac{\Gamma{(p+\frac{1}{2})}\rho^{q}
(1-\rho)^{p-q}}{\Gamma{(q+1)}\Gamma{(p-q+1)}\Gamma{(\frac{1}{2})}}\right)\left(\frac{1}{pd-q-d+1}\right)\alpha^{pd-q}\\&-\sum_{n=0}^{\infty}\sum_{k=0}^{n}
\sum_{p=0}^{\infty}\sum_{q=0}^{p}\Biggl\{\frac{\sqrt{\pi}}{(2d-2)}\left(\frac{\Gamma{(n+\frac{1}{2})}\Gamma{(p+\frac{1}{2})}\rho^{k+q}
(1-\rho)^{n+p-k-q}}{\Gamma{(k+1)}\Gamma{(q+1)}
\Gamma{(n-k+1)}\Gamma{(p-q+1)}\Gamma{(\frac{1}{2})}\Gamma{(\frac{1}{2})}}\right)\\&\times\frac{\Gamma \left(\frac{nd+pd-k-q+1}{2 d-2}\right)}{\Gamma \left(\frac{nd+pd-k-q+1}{2 d-2}+\frac{1}{2}\right)}\left(
\frac{1}{pd-q-d+1}\right)\alpha^{nd+pd-k-q}\Biggr\},
\end{split}
\end{equation}
where $\epsilon$ is the UV regulator in the boundary theory. The multiple infinite sum appearing in the above result is hard to compute exactly for arbitrary values of the set of parameters present in the theory. Also one needs to check the convergence of the infinite sums present in \ref{I22}. Hence in the following subsections, we have considered the low and high effective temperature limit on the volume and correspondingly evaluated the integrals and verified the convergence of the infinite sums.

\subsection{Volume complexity at low effective temperature}

In the low temperature limit $(\alpha\rightarrow 0)$, one can terminate the infinite sum appearing in the above expression for the volume keeping terms upto $d$-th power in $\alpha$ so that the final result for the volume at low temperature is given as,
\begin{equation}
\begin{split}
\label{Vlow}
V&=\frac{L^{d-2}lR^{d}}{(d-1)}\frac{1}{\epsilon^{d-1}}+R^{d}\left(\frac{L}{l}\right)^{d-2}\frac{\rho}{2} \left(\frac{4 \pi l T_{f}}{d}\right)^{d-1}\log{\left(\frac{\epsilon}{l}\right)}\\&+R^d\left(\frac{L}{l}\right)^{d-2} \Biggl\{\mathcal{V}_{0}+\mathcal{V}_{1}\rho\left(\frac{4 \pi l T_{f}}{d}\right)^{d-1}+\mathcal{V}_{2}(1-\rho )\left(\frac{4 \pi l T_{f}}{d}\right)^d+\mathcal{O}\left(T_{f}l\right)^{2d-2}\Biggr\},
\end{split}
\end{equation}
where, $\mathcal{V}_{0/1/2}$ are $\mathcal{O}(1)$ constants and depends only on the dimension $d$. The explicit forms are given in Appendix-{\bf {A}}. Again the leading order correction term is proportional to the quark density $\rho$ and it also varies as the volume of the entangling hypersurface as expected.
In the above result at low temperature, we get a logarithmic UV divergent term with a $\rho$ dependent coefficient apart from the usual power law divergence at the boundary $z\rightarrow 0$. Also notice that the logarithmic divergence is devoid of any dependance on the dimension $d$ which indicates that this divergence will be present irrespective of the dimension of the space time. It is very important to note that the $\rho$ dependent terms appeared so far in various expressions of HEE and EWCS are basically non-trivial corrections of the quantum entanglement arising in the thermal plasma due to the presence of the back reaction we have considered. However, by looking at the logarithmic term in the expression of low temperature limit of the subregion volume complexity, one may appreciate that this terms is actually a universal term combining both UV cut-off and also the IR cut-off of the boundary theory. By IR cut-off we mean the low temperature/large length scale defined by $l$. \footnote{This comment we appended here came to our notice after a communication with Aron Wall.}

\subsection{Volume complexity at high effective temperature}
The high temperature limit is effected as usual by the limit $z_{t}\rightarrow z_{H}$ and we find that the subregion volume complexity diverges. This diverging result is appearing from the large values of the indices denoted as $n$ and $p$ contained in the multiple sums in $I_{1}$ and $I_{2}$ as given in (\ref{I11}) and (\ref{I22}). It is more useful to express $I_{1}$ by not using the infinite series expansion of the term $(1/\sqrt{h})$,
\begin{equation}
\begin{split}\label{I19}
I_{1}=\frac{l}{2}\left(\frac{z_{t}^{d-2}}{d-1}\frac{1}{\epsilon^{d-1}}+\frac{\rho}{2}\alpha^{d-2}\log{\left(\frac{\epsilon}{z_{t}}\right)}\right)
-\frac{l}{2z_{t}}\left(\frac{1}{d-1}\right)+\frac{l}{2z_{t}}\mathcal{Y}(\alpha,d,\rho)+\frac{l}{2z_{t}}\int_{0}^{1}\frac{du}{\sqrt{h}},
\end{split}
\end{equation}
where we have denoted $\mathcal{Y}(\alpha,d,\rho)$ as the following integral,
\begin{equation}\label{Y}
\mathcal{Y}(\alpha,d,\rho)=\int_{0}^{1}du\left(\frac{1}{u^{d}\sqrt{h}}-\frac{1}{u^{d}}-\frac{\rho}{2u}\alpha^{d-1}
-\frac{1}{\sqrt{h}}\right).
\end{equation}
Notice that in (\ref{I19}) we have added and subtracted a term $\frac{l}{2z_{t}}\int_{0}^{1}\frac{du}{\sqrt{h}}$ in order to make the definite integral $\mathcal{Y}(\alpha,d,\rho)$ finite in the high temperature limit $(\alpha\rightarrow 1)$. As an example, in the following we list a few values of $\mathcal{Y}(\alpha=1,d,\rho)$ in $d=4$ as,
\begin{equation}
\mathcal{Y}(d=4,\rho=1)=-1.004,~~~\mathcal{Y}(d=4,\rho=2)=-1.422,~~~\mathcal{Y}(d=4,\rho=3)=-1.770
\end{equation}
The final result for $I_{1}$ in the high effective temperature limit is given as,
\begin{equation}
\begin{split}\label{I1}
I_{1}=\frac{l}{2}\left(\frac{z_{H}^{d-2}}{d-1}\frac{1}{\epsilon^{d-1}}+\frac{\rho}{2}\log{\left(\frac{\epsilon}{z_{H}}\right)}\right)
-\frac{l}{2z_{H}}\left(\frac{1}{d-1}\right)+\frac{l}{2z_{H}}\mathcal{Y}(\alpha,d,\rho)+\frac{l}{2z_{H}}\int_{0}^{1}\frac{du}{\sqrt{h}}.
\end{split}
\end{equation}
Now, turning our attention to $I_{2}$, we note that, unlike $I_1$ in this case we encounter double integration in terms of the variables $u$ and $u'$ respectively. In contrary to the choice of methodology we followed in evaluating $I_{1}$, here to evaluate the $u^{\prime}$ integration in $I_{2}$ as given in \ref{complex}, we must use series expansion of $1/\sqrt{h(u^{\prime})}$,
\begin{equation}
\begin{split}
I_{2}&=\sum_{p=0}^{\infty}\sum_{q=0}^{p}\frac{\Gamma{(p+\frac{1}{2})}\rho^{q}(1-\rho)^{p-q}}
{\Gamma{(q+1)}\Gamma{(p-q+1)}\Gamma{\frac{1}{2}}}\left(\frac{1}{pd-q+d}\right)\alpha^{pd-q}\\&\times\int_{0}^{1}\frac{du}{\sqrt{h(u)}}
u^{pd-q}~~{}_2 F_1\left(\frac{1}{2}, \frac{pd-q+d}{2d-2},\frac{pd-q+d}{2d-2}+1; u^{2d-2}\right).
\end{split}
\label{I2new}
\end{equation}
In the high temperature limit, $\alpha\rightarrow 1$, after computing the $u'$ integration of $I_2$ as given in (\ref{I2new}) the value of the upper limit of $u$ variable of the existing integral of $I_2$, i.e $u=1$ sources possible divergence in the final expression. So to figure out the divergent terms we make a series expansion of the integrand in the above expression of $I_2$ near $u=1$ and get the following,
\begin{equation}
\begin{split}
I_{2}=\left\{\frac{l}{2z_{t}}\int_{0}^{1}\frac{du}{\sqrt{h}}-\sqrt{\frac{2}{d-1}}\frac{1}{\sqrt{\delta}}
\sum_{p=0}^{\infty}\sum_{q=0}^{p}\frac{\Gamma{(p+\frac{1}{2})}(d-\delta)^{q}(1-d+\delta)^{p-q}}
{\Gamma{(q+1)}\Gamma{(p-q+1)}\Gamma{\left(\frac{1}{2}\right)}}+\textrm{Finite}~~ \textrm{terms}\right\},
\end{split}
\label{l2new1}
\end{equation}
where in the expression the first two terms are the only divergent pieces in $I_2$. Note that we have replaced $\rho$ by $(d-\delta)$ in the above expression such that $\delta$ is a non-zero positive number.
It is also important to note that the divergent terms contained in (\ref{l2new1}) correspond to only $\delta \ne 0$, whereas analogue expression for such divergence corresponding to $\delta = 0$ has been separately mentioned later in (\ref{I2new2}).
In the above equation (\ref{l2new1}), the first term cancels exactly with the final term of (\ref{I1}) which was initially considered to make equation (\ref{Y}) finite in the limit $\alpha \to 1$. The other divergent term in (\ref{l2new1}) with double sum turns out to be proportional to the infinite part of the width $l$ which can be easily verified from equation (\ref{width}) by performing a series expansion of the integrand near $u=1$, namely,
\begin{equation}
\begin{split}
\Biggl(\frac{l}{2z_{H}}\Biggr)_{\textrm{infinite}}=\left(\sqrt{\frac{2}{d-1}}\right)
\sum_{p=0}^{\infty}\sum_{q=0}^{p}\frac{\Gamma{(p+\frac{1}{2})}(d-\delta)^{q}(1-d+\delta)^{p-q}}
{\Gamma{(q+1)}\Gamma{(p-q+1)}\Gamma{\left(\frac{1}{2}\right)}}
\end{split}
\end{equation}
So in the final result for the divergent part of $I_{2}$ is given as,
\begin{equation}
\Biggl(I_{2}^{(\rho=d-\delta)}\Biggr)_{\textrm{infinite}}=\frac{l}{2z_{H}}\int_{0}^{1}\frac{du}{\sqrt{h}}-\frac{1}{\sqrt{\delta}}\Biggl(\frac{l}{2z_{H}}\Biggr)_{\textrm{infinite}}.
\label{inf1}
\end{equation}
Finally, for $\rho=d$ (or equivalently $\delta=0$), the diverging terms in $I_2$ can be obtained as,
\begin{equation}
\begin{split}
\Biggl(I_{2}^{(\rho=d)}\Biggr)_{\textrm{infinite}}=\frac{l}{2z_{H}}\int_{0}^{1}\frac{du}{\sqrt{h}}+\sqrt{\frac{2d}{d-1}}\frac{l}{2z_{H}}+\left(-\frac{2\sqrt{2}}{d\sqrt{d-1}}\right)
\Biggl(\frac{l}{2z_{H}}\Biggr)_{\textrm{infinite}}.
\end{split}
\label{I2new2}
\end{equation}
Again the first term in (\ref{I2new2}) cancels the last term in (\ref{I1}) and we are left with two infinite terms expressed in terms of the width $l$ as shown in the above result. The whole point of the above analysis is to express the diverging terms appearing in the IR scale (large length scale) by an intrinsic parameter of the boundary theory which in this case is given by the width $l$ of the entangling surface.
\section{Conclusion}
\begin{figure}

 \centering
  \includegraphics[width=.5\linewidth]{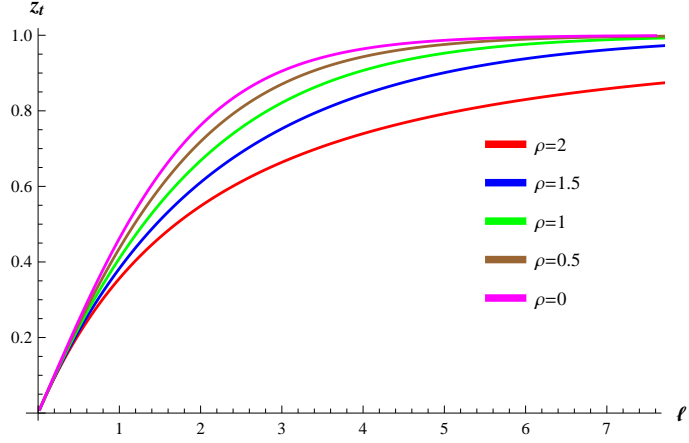}
  \caption{Plot of $z_{t}$ vs $l$ for different values of $\rho$}
  \label{sub44}
\end{figure}

In this paper our primary motivation is to find out the effects of finite quark density in the boundary theory on three different measures related to quantum information theory. After doing this analysis, we come up with the following observations.
\begin{itemize}
\item The first measure is the computation of entanglement entropy at low and high effective temperature using the prescription by Ryu and Takayanagi. We have explicitly observed from (\ref{highS}) and (\ref{rhodiff}) that the EE shows increasing behavior with quark density for a given temperature. The same behavior can also be realized from the plot of EE with respect to $l$ for different values of $\rho$ in figure-(\ref{sub45}).

\item Similar to EE, an increasing behavior of EWCS with the quark density has also been obtained and can be realized graphically from figure-\ref{fig4}. Regarding the computation of entanglement wedge cross section in the presence of any kind of  back reaction has not been reported elsewhere prior to our present work.

\item Finally, the low temperature result (\ref{Vlow}) for subregion volume complexity, in particular the leading order correction term proportional to the quark density $\rho$, suggests the complexity for the subregion increases with $\rho$ for a given temperature.

\item Interestingly, the behavior for all the above three measures with respect to the quark density can be realized from the profile of the entangling surface into the bulk geometry. In figure-(\ref{sub44}), we have plotted the turning point $z_{t}$ as a function of the strip width $l$ for different values of the quark density $\rho$. One can observe that the RT surface extends deep into the bulk and reach near the horizon for smaller values of $l$ when the corresponding quark density $\rho$ is relatively lower. On the other hand, for higher values of $\rho$ this extension into the bulk is relatively slower with respect to the width $l$. As a result the area of the RT surface as well as the volume enclosed by it will be larger for higher $\rho$. Meaning, both EE and volume complexity increases with the quark density. Regarding the behavior of EWCS, referring to figure-(\ref{sub111}), the RT surface $\gamma_{D}$ for the separation region with width $D$, the turning point for low quark density denoted by $z_{t}^{\textrm{small}~\rho}(D)$ will be greater than the corresponding value for higher density $z_{t}^{\textrm{large}~\rho}(D)$. So for the phase transition to take place, separation between the two subregions needs to take higher value as the corresponding quark density becomes large. Hence the EWCS also increases with the quark density $\rho$.

\item By considering the zero quark density limit in the final expressions of all the three measures we could successfully reproduce the corresponding result for AdS-BH background. In this regard it is important to keep the sub-leading correction terms also in order to get the exact AdS-BH results.
    \item In the computation of subregion volume complexity we have obtained a log divergent term with coefficient which is proportional to the quark density of the boundary theory. So the occurrence of this term is only because of the finite back reaction of the finite heavy quark to the super Yang-Mills thermal plasma that we considered here. It is also important to note that the argument of this logarithm is a dimensionless ratio between the UV cut-off $\epsilon$ and the strip width $l$, where $l$ is some IR cut-off of the theory. Hence the coefficient of both $\log{(\epsilon)}$ which is a UV divergence, and $\log{(l)}$ which on the other hand is associated with some IR aspects of the theory, has exactly the same coefficient. In that sense the coefficient of the log term carries information about the system at all energy scale and not just the UV regime.
        \end{itemize}
Before closing, we would like to mention that if one could find out the quantitative change in the boundary theory due to the presence of back reaction even in a special case as  $d=4$ large $N$ $\mathcal{N} = 4$ super Yang-Mills thermal plasma, it would be very interesting to verify our results from the CFT perspectives. In particular, to understand the field theoretic origin of realizing the logarithmic universal term appearing in the result of subregion complexity would be very interesting. We leave this way of performing field theoretic analysis for our future work.
As mentioned in the introduction, we hope that the universal features of various entanglement measures and subregion volume complexity would be very important to understand the quantum correlation in a realistic system such as quark-gluon plasma.
\section*{Acknowledgements}
We would like to thank Aron Wall, Komeil Babaei Velni, Arjun Bagchi and Rajesh Kumar Gupta for useful discussions. SC is partially supported by ISIRD grant 9-252/2016/IITRPR/708. SP is supported by a Junior Research Fellowship from the Ministry of Human Resource and Development, Government of India. KS is supported by the institute post doctoral fellowship from IIT Ropar, Ropar.

\appendix
\section{Subregion volume complexity at low effective temperature.}
The $d$ dependent constant terms $\mathcal{V}_{0}, \mathcal{V}_{1}, \mathcal{V}_{2}$ as appearing in the expression for the co-dimension one volume of the RT hypersurface at low effective temperature (\ref{Vlow}) are given below,
\begin{equation}
\mathcal{V}_0=\frac{2^{1-d}\pi^{\frac{1}{2}-\frac{d}{2}}\biggl(\frac{\Gamma{[\frac{1}{-2+2d}]}}{\Gamma{[\frac{d}{-2+2d}]}}\biggr)^{-1+d}\biggl\{2\pi\Gamma{[\frac{d}{-2+2d}]^2}-(-1+d+2\pi)\Gamma{[\frac{1}{-2+2d}]}\Gamma{[\frac{1}{2}+\frac{d}{-2+2d}]}\biggr\}}{(-1+d)^2\Gamma{[\frac{1}{-2+2d}]}\Gamma{[\frac{1}{2}+\frac{d}{-2+2d}]}}
\end{equation}

\begin{equation}
\begin{split}
\mathcal{V}_1=\frac{1}{2}\biggl\{-&\frac{16(-2+d)\pi^3\Gamma{[\frac{d}{-2+2d}]^4}}{d(-1+d)^3\Gamma{[\frac{1}{-2+2d}]^3}\Gamma{[\frac{1}{2}+\frac{d}{-2+2d}]}}+\frac{2\pi\Gamma{[\frac{d}{-2+2d}]}\left(\Gamma{[\frac{1-2d}{-2+2d}]}-\Gamma{[\frac{2-3d}{-2+2d}]^2}\Gamma{[\frac{d}{-2+2d}]}\right)}{(-1+d)^2\Gamma{[\frac{2-3d}{2-2d}]}\Gamma{[\frac{1}{-2+2d}]}\Gamma{[\frac{1}{2}+\frac{d}{-2+2d}]}}\\& +\log{\biggl[\frac{\Gamma{[\frac{1}{-2+2d}]}}{2\sqrt{\pi}\Gamma{[\frac{d}{-2+2d}]}}}\biggr]+\biggl[\pi \Gamma{[\frac{d}{-2+2d}]^2}  \biggl(\frac{ 8(1+d^2+2d(-1+\pi)-4\pi)\pi\Gamma{[\frac{1-2d}{2-2d}]}}{d(-1+d)^3\Gamma{[\frac{1-2d}{2-2d}]}\Gamma{[\frac{1}{-2+2d}]}}\\&-\frac{d(-1+d)\Gamma{[\frac{1}{-2+2d}]}\left\{ \psi_{(0)}[\frac{1-2d}{2-2d}]- \psi^{(0)}[\frac{d}{2-2d}]\right\}}{d(-1+d)^3\Gamma{[\frac{1-2d}{2-2d}]}\Gamma{[\frac{1}{-2+2d}]}}\biggr)\biggr]\biggr\}
\end{split}
\end{equation}

\begin{equation}
\begin{split}
\mathcal{V}_2= \frac{\sqrt{\pi}\Gamma{[\frac{d}{2(-1+d)}]}}{\Gamma{[\frac{1}{2(-1+d)}]}}&\biggl\{ \Gamma{[\frac{1}{-2+2d}]^2}+\frac{\frac{2\pi}{(-1+d)^2}\Gamma{[\frac{d}{-2+2d}]}\Gamma{[\frac{1}{-2+2d}]}}{(1+d)\Gamma{[\frac{1+d}{-2+2d}]}\Gamma{[\frac{1}{2}-\frac{d}{-1+d}]}}\biggl[\frac{(-2+d)\Gamma{[\frac{1+d}{-2+2d}]}}{\Gamma{[\frac{d}{-1+d}]}}\\&+\Gamma{[\frac{d}{-1+d}]}\biggl((1+d)\Gamma{[\frac{1+d}{-2+2d}]}+2(1+d^2+4\pi-2d(1+\pi))\Gamma{[\frac{1}{2}-\frac{d}{-1+d}]}\biggr)\\&-\frac{(-1+d)\Gamma{[\frac{d}{-2+2d}]}}{\Gamma{[\frac{1}{2}-\frac{d}{-1+d}]}}\biggr]+\frac{8(-2+d)\pi^2\Gamma{[\frac{d}{-2+2d}]^3}\Gamma{[\frac{d}{-1+d}]}}{(-1+d)^2(1+d)\Gamma{[\frac{1+d}{-2+2d}]}\Gamma{[\frac{1}{2}+\frac{d}{-1+d}]}}\biggr\}
\end{split}
\end{equation}

\end{document}